\documentclass[letterpaper,12pt]{article} 
\usepackage{osajnl2} 
\usepackage[draft]{hyperref}

\begin{document}

\title{A simple optical system for interpreting coherence theory.}

\author{Damien P. Kelly$^{ 1,2}$}

\address{$^1$ Institut f$\ddot{u}$r Mikro-und Nanotechnologien, \\Macro-Nano, \\Fachgebiet Optik Design, \\Technische Universit$\ddot{a}$t Ilmenau, \\ P.O. Box 10 05 65, \\ 98684 Ilmenau,\\Germany.}

\address{$^2$ Oryx Optics Ltd, \\19 The Elms, \\Athlone, \\Ireland.}

\email{damienpkelly@gmail.com} 

\begin{abstract}
A new theoretical technique for understanding, analyzing and developing optical systems is presented. The approach is statistical in nature, where information about an object under investigation is discovered, by examining deviations from a known reference statistical distribution.  A Fourier optics framework and a scalar description of the propagation of monochromatic light is initially assumed.  An object (belonging to a known class of objects) is illuminated with a speckle field and the intensity of the resulting scattered optical field is detected at a series of spatial locations by point square law detectors.  A new speckle field is generated (with a new diffuser) and the object is again illuminated and the intensities are again measured and noted.  By making a large number of these statistical measurements - an ensemble averaging process (which in general can be a temporal or a spatial averaging process) - it is possible to determine the statistical relationship between the intensities detected in different locations with a known statistical certainty.  It is shown how this physical property of the optical system can be used to identify different classes of objects (discrete objects or objects that vary continuously as a function of several physical parameters) with a specific statistical certainty.  The derived statistical relationship is a fourth order statistical process and is related to the mutual intensity of scattered light.  A connection between the integration time of the intensity detectors and temporal coherence and hence temporal bandwidth of a quasi-monochromatic light source is discussed. We extend these results to multiple wavelengths and polarizations of light. We briefly discuss how the results may be extended to non-paraxial optical systems.\\  
\\
\end{abstract}

\maketitle
\newpage
%
%
\section{Introduction}
In this paper we develop a simple `thought-model' to help us understand coherence theory for scalar optical fields. This model will allow us to consider light with different coherence and polarization states and which may contain many different contributing wavelengths. We will examine how this light field interacts with some object and then how the resulting scattered light is detected; specifically, we will examine point intensity detectors that measure the intensity of the scattered light over a finite aquisition time. We will examine this problem using a Fourier optics framework.  A particular advantage of Fourier optics is that the relative simplicity of the treatment allows one to develop intuitive models that provide significant insight into the behaviour of optical systems \cite{GO1966, LohmannBook2006, KellyFiniteAp3D,Kelly:07b}.
\\
\\
Here, we assume that a scalar model of light propagation is valid; that there are no current sources or charges in our medium; and hence that each vectorial component of the electromagnetic field can be treated independently from the other components \cite{GO1966}.  This implies that the different vectorial components do not interact with or `see' each other.  Within the scalar description of light propagation, it is common to use diffraction integrals to relate the complex scalar field in one plane, to the diffracted field, in an axially displaced plane.  Non-paraxial diffraction integrals such as the Rayleigh-Sommerfeld I and II or the Kirchoff can be derived from the time-independent Helmholtz equation and strictly speaking are correct only for propagation through a uniform medium with constant refractive index.  While these integrals are exact solutions for the scalar wave equation, the analysis is easier still when the paraxial approximation is assumed. Now we can describe light propagation using the Fresnel transform. A second and necessary part of our analysis is to model the interaction of scalar field with some object. We imagine that if a field is incident on the object, then the field `immediately after' the object is calculated by assuming the `Thin Element Approximation' (TEA). The real physical dimensions of the object are thus ignored, the object is imagined to be infinitely thin, and that the field before and after the object are related to each other by multiplying the incident field with a transmittance function. This type of approach is discussed at length by Goodman \cite{GO1966}, and will be adopted in the forthcoming analysis.  These approximations are commonly used to analyze speckle theory, aspects of coherence theory and statistical optics, see for example \cite{Good:1985, mandel1995optical, Good:07,wolf2007introduction}. 
   \begin{figure}[htp!]
   \label{CSpaperFigure1}
     \includegraphics[height=9cm]{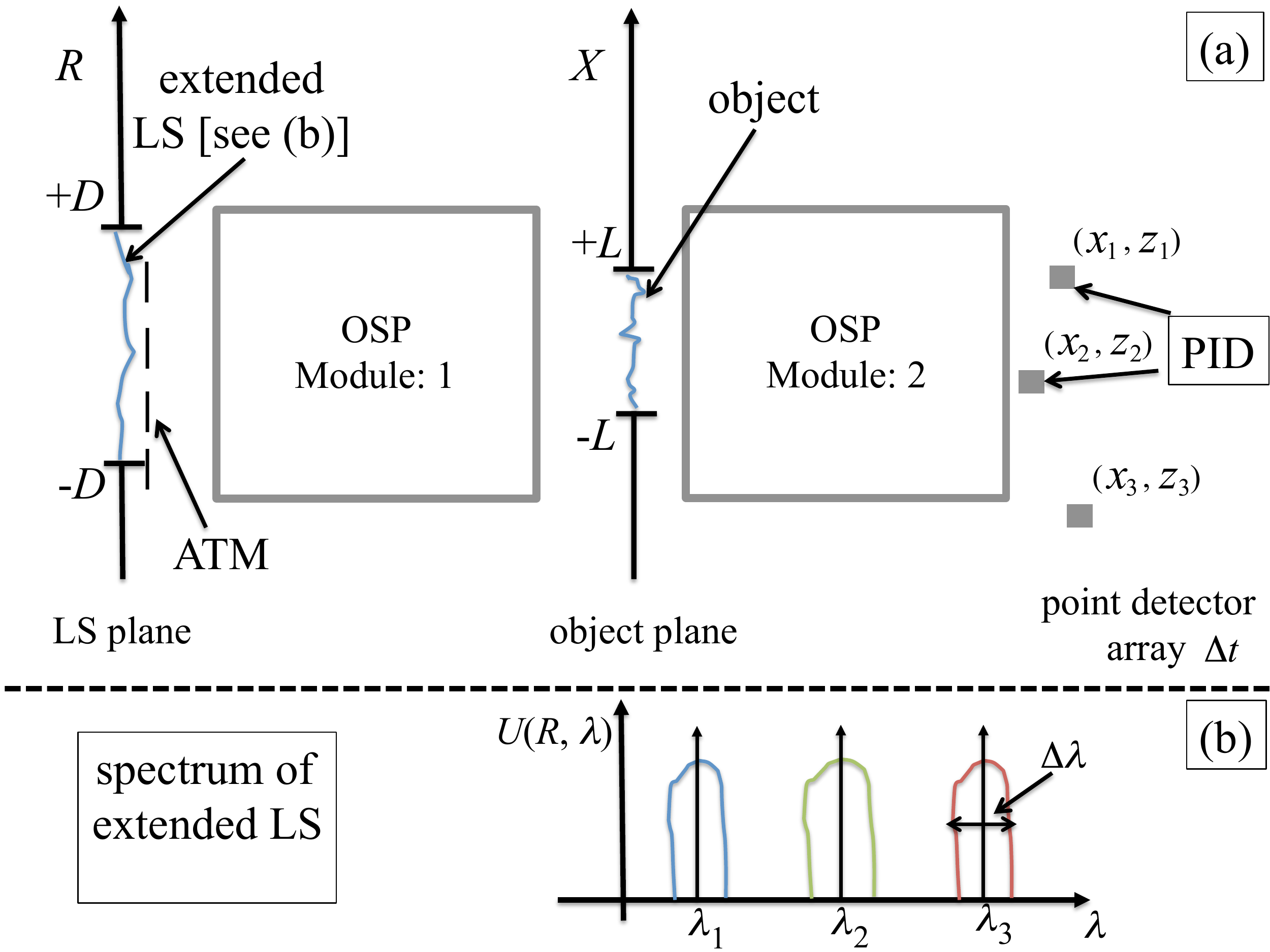}
     \caption{(a) Schematic of the type of general optical system we are examining. The Light Source (LS) we are considering is a multi-wavelength extended source, where the Amplitude Transmittance Mask (ATM) can be used to select areas of light source that are used for the illumination of the sample. The light from this source is modified by OSP Module 1 before it illuminates a sample located in the object plane. The resulting scattered light is then collected by OSP Module 2 before it is measured by an array of Point Intensity Detectors (PID) that are both wavelength and polarization sensitive. The acquisition time (integration time) of the detectors is $\Delta t$. Inset (b) depicts the spectrum of this extended LS. }
    \end{figure}
\\
\\
In Fig. \ref{CSpaperFigure1} we sketch the general type of optical system we would like to analyze. It consists of an extended quasi-monochromatic light source, whose spatial coherence properties in the source plane are totally incoherent, (i.e. the mutual coherence function is delta correlated, see Eq. (5.6.2) in Ref. \cite{Good:1985}). We further allow that there may be several discrete quasi-monochromatic wavelengths  with orthogonal polarizations illuminating the sample simultaneously; we will use different transmittance masks to select specific parts of the extended source to perform the illumination. The light emanating from this source is then subject to a signal processing operation - by Optical Signal Processing (OSP) Module 1.  These OSP modules consist of combinations of lenses and sections of free space and can be described using an LCT (see Section 1 of Ref. \cite{Kelly2011} or \cite{Yura:99}). This light then illuminates a sample that we are interested in examining that is situated in the object plane. The interaction of the light and the sample is modelled using the TEA. Then the resulting scattered light is once again subject to a signal processing operation - OSP Module 2 - before its intensity is detected by an array intensity detectors. These intensity detectors will, in general, be color and polarization selective so that multi-wavelength measurements can be made simultaneously on different polarizations of the scattered and processed light. We will be interested in relating the acquisition time of the intensity detectors, $\Delta t$, to coherence length of the individual discrete quasi-monochromatic illuminating light sources and we note that the coherence length of the light is usually related to the inverse of its spectral width and so to $\Delta \lambda$ in Inset (b) of Fig. \ref{CSpaperFigure1}, see for example Chap. 5 in Ref. \cite{Good:1985}.
\\
\\
In general this type of optical system is difficult to analyze due to all the different aspects that need to be considered. Instead we propose to develop a simpler analytical model or `Gedankenexperiment'  that may be used as a tool to understand how different aspects of coherence theory, object detection, signal processing and electronic detection interact with each other. An optical system that could fulfill this role is depicted in Fig. \ref{CSpaperFigure2}. There are several differences between this optical system and the more general one depicted in Fig. \ref{CSpaperFigure1}. First this simpler system uses an optical Fourier transform to process the illuminating light before it is incident on the object, and then a Fresnel transform is used to model the propagation of the light after the object to the locations of the intensity detector array. Secondly, we note that in this system the light we use is now considered to be a perfectly mono-chromatic and spatially coherent plane wave which is incident on a diffuser pair. This diffuser pair produces a Random Speckle Field (RSF) which serves as an instance of the random interrogating signal or illumination field. The statistical properties of such a field have been examined by many different authors \cite{Dainty-76,Good:1985, Good:07, Yura:87, Yura:93,Yura:99,Rose:98,Yoshimura:92,Yoshimura:93}. A speckle field follows well known statistical distributions for its intensity and phase distributions. 
\\
\\
Suppose we move one of these diffusers, it is possible to generate a new and statistically independent field. While this second speckle field is completely different from the first field it will have identical statistical properties. Other authors have examined the time dependent statistical properties as these diffusers are moved relative to each other \cite{Li:11,Li:11b,Li:12,Li:13,Li:13b}. It is possible to determine a definite statistical relationship between the intensities at different locations. Alternative methods for producing a similar effect to the diffuser pair illumination are discussed here \cite{Cabezas:15}.
\\
\\
Having established these known statistical properties for a speckle field, we now consider what happens to this known statistical distribution when an object (which we think of as a symbol or message) is inserted into the optical path. It is expected that the optical field interacts with the sample and modifies the statistical properties of the scattered field \cite{Takeda:05,Naik:10}.  Hence by measuring changes in the resulting statistical distribution we can identify the object under investigation. We can perform this measurement in some cases using a single speckle field for illumination and by making a series of spatial averages - spatial averaging. Alternatively we can illuminate the object with a sequence of random speckle fields and perform a time averaging operation as is discussed here \cite{Li:11,Li:11b,Schurig2015-2}. 
\\
\\
In the following section we derive a mathematical description of the optical system in Fig. \ref{CSpaperFigure2}, deriving the relevant equations that follow from the assumptions we have just outlined. In Section 3, we examine the characteristics of the resulting equations and derive a correlation function - that is very similar to the mutual coherence function of discussed by Goodman in Ref. \cite{Good:1985}. Our correlation function however also depends on the nature of the ATM in Fig. \ref{CSpaperFigure2}. Some specific calculated examples are presented for a range of different objects. And we shall see that the analysis naturally extends to include partially coherent effects, including the relationship between $\Delta t$ and $\Delta \lambda$ identified in Fig. \ref{CSpaperFigure1}. In Section 4, we link the accuracy of our mathematical model to practical experimental measurements. In Section 5, we discuss how these results can be extended and used to identify objects from particular types of classes of objects. We discuss two forms of this problem: (i) Where the objects come from a finite set of unique discrete objects; for example the set consisting of either a lens or a grating, and (ii) Where the objects are described using a basis set and can change continuously as a function of several physical parameters; this could be a lens with a continually varying focal length.  We find that a-priori information about the nature of the object is essential so that different objects can be distinguished from each other. In Section 6, we discuss how the scalar analysis presented can be extended to include polarization effects and different wavelengths of light. These results are exact within the assumptions we have made and in the conclusion we briefly discuss how these could be extended to other models of optical propagation and to models of the interaction of light and material.
\section{The optical system}
\begin{figure}[htp!]
   \label{CSpaperFigure2}
     \includegraphics[height=10cm]{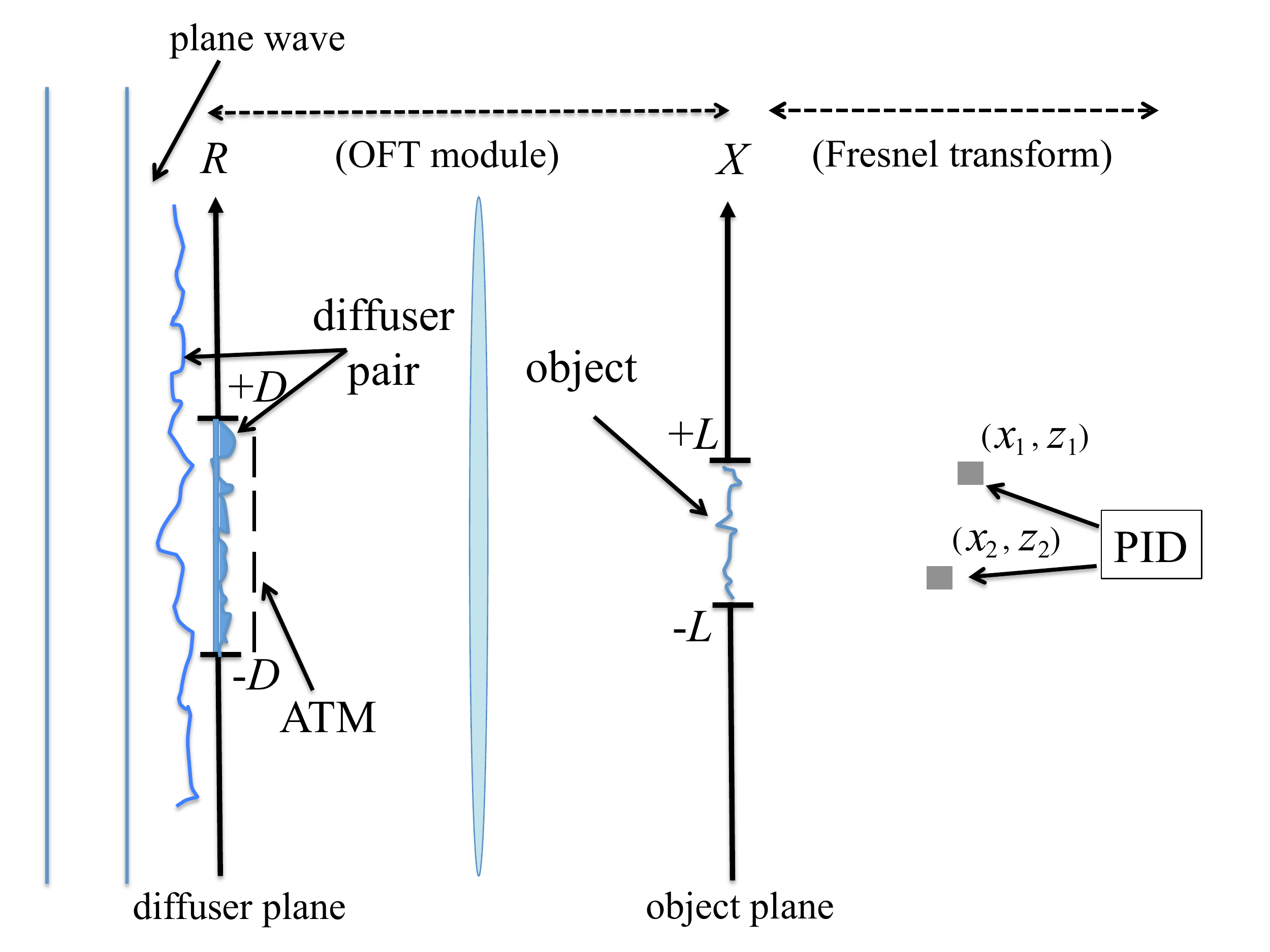}
     \caption{Optical setup for generating a speckle field.  A plane wave is incident on a diffuser and the phase of the field is randomized. The field behind the diffuser is a Random Speckle Field (RSF). This field is Fourier transformed by a K$\ddot{\mathrm{o}}$hler lens and illuminates the target located in the $X$ plane. The resulting scattered intensity is measured at locations $P_1$ and $P_2$ by Point Intensity Detectors (PID). Inset (i) depicts the zoomed in section A.  To calculate the complex amplitude at $P_3$ one must consider the contribution at each point (several are depicted) in the diffuser plane.  Inset (ii) is the zoomed in section B.  The diffuser consists of two Rough Surfaces (RS).  RS1 can rotate or translate relative to RS2 to produce statistically independent speckle fields. Points $P_1$ and $P_2$ are spatially located at $(x_1, z_1)$ and $(x_2, z_2)$ respectively.}
    \end{figure}

Consider the optical system depicted graphically  in Fig. \ref{CSpaperFigure2}.  A perfectly collimated plane wave (of unit amplitude) is incident on a doubly scattering diffuser-pair.  The extent of this diffuser-pair is limited by a hard aperture with a width of $2D$. Each diffuser is supposed to be optically rough imparting phase changes that are uniformly distributed and greater than 2$\pi$. The surface profiles of the diffusers are modeled with two separate continuous functions. Since these two surface profile functions describe two different diffusers it is reasonable to assume that they are statistically independent; however they can still have identical statistical properties and this is now assumed. For example the width of the auto-correlation of each surface profile would be the same. We further assume that immediately after the diffuser-pair, only the phase of the incident plane wave is modified while the amplitude remains constant. We thus write an expression for the field `just after' the diffuser-pair as 

\begin{eqnarray}
\label{S2.1}
V\left(R \right) = \exp \left[ j \Phi_n \left(R \right) \right] p_D \left(R \right),
\end{eqnarray}
where $\Phi_n \left(R \right)$ is a random function which changes the phase of $V(R)$ for each position of $R$. We note that $j = \sqrt{-1}$, and that the aperture function is defined in the following manner
\begin{eqnarray}
\label{S2.2}
p_D \left(R \right) = \left \{ \begin{array}{ll} 1, &  \mathrm{when}\  |R| < D   \\
					   			0, & \mathrm{ otherwise.}\\
					   \end{array}\right. 
\end{eqnarray}
In preparation for later discussion we briefly consider the characteristics of this diffuser-pair.  We note that if one diffuser is moved relative to the other even by a very small amount - on the order of the auto-correlation width of the diffuser surface profile - it will generate a new field $V(R)$ that is statistically independent from the previous field, see \cite{Shirley:88,Shirley:89},  also Chapter 3 and 6 from Ref. \cite{Good:07}, and \cite{ODonnell:82, Newman:85,Yoshimura:92, Li:13, Li:13b}.  A rough estimate for the amount of relative translation between the two diffusers is about 10 $\mu$m \cite{Schurig2015-2}. Hence for two different relative positions of the diffusers we get two different phase functions, $\Phi_1 \left(R \right)$ and $\Phi_2 \left(R \right)$ which have identical statistical properties but are statistically independent of each other. In practice it is possible to generate a very large number of these statistically independent phase functions by moving the diffusers relative to each other and this fact is reflected in our notation where the subscript `n' in $\Phi_n \left(R \right)$ indicates a particular instance of a random phase function. We now briefly consider the power of $V(R)$, which can be calculated according to 
\begin{eqnarray}
\label{S2.2b}
P_{DP} & = & \int_{-\infty}^\infty V(R)V^*(R) dR \nonumber\\
	     & = & 2D.
\end{eqnarray}
A spatial filtering operation is then implemented on  on $V(R)$ with a pinhole mask, see Fig. \ref{CSpaperFigure2}, and hence only a portion of $V(R)$ is allowed to propagate into the optical system. Initially this pinhole mask will consist only of a single pinhole, however later in the text we will generalize this expression by adding more pinholes to the mask and then extending the analysis so that a continuum of point sources can be considered. Generally then we use an Amplitude Transmittance Mask (ATM) to select appropriate areas of the source to illuminate the object. This first part of the optical system, the illuminating optics, is modeled as an ideal Fourier transforming system \cite{GO1966}, where the lens is assumed to be both `thin' and infinite in extent \cite{KellyFiniteAp2D,KellyFiniteAp3D,Kelly:07b}. For incoherent light sources this configuration is sometimes referred to as
K$\ddot{\mathrm{o}}$hler illumination. We now write an expression for the resulting field that illuminates the target situated in the object plane,
\begin{eqnarray}
\label{S2.3}
U\left(X \right) & = & K_f \int_{-\infty}^\infty V(R) \delta \left(R-\Delta_1 \right) \exp \left(\frac{ - j 2 \pi R X }{\lambda f} \right) dR \nonumber\\
& = & K_f \exp \left(j \phi_1 \right) \exp\left(\frac{ - j 2 \pi \Delta_1 X }{\lambda f} \right) 
\end{eqnarray}
where the constant $K_f = 1/\sqrt{j \lambda f}$ will be dropped for notational convenience from here on. The pinhole in Fig. \ref{CSpaperFigure2} is modeled as a Dirac delta point source located at $R = \Delta_1$ in the diffuser plane. The wrapped phase value, $\phi_1^n$ = $\Phi_n(\Delta_1)$, is a uniformly distributed random variable lying between 0 and $2 \pi$. If we were to move the diffuser-pair relative to each other, then a new value of $\phi_1^n$ would be generated, which would again be a uniformly distributed random variable lying between 0 and $2 \pi$. Again for notational simplicity we will suppress the `n' superscript and refer simply to $\phi_1$.We recognize that Eq. (\ref{S2.3}) is a propagating plane wave with a random phase, i. e.
\begin{eqnarray}
\label{S2.4}
U\left(X \right) & = & \exp \left(j \phi_1 \right) \exp\left( - j 2 \pi f_{x1} X \right),
\end{eqnarray}
with $f_{x1} = \Delta_1/\left(\lambda f \right)$.
\\
\\
Let us describe the target in the object plane as $O(X)$ such that 
\begin{eqnarray}
\label{S2.5}
\tilde{O} \left(X \right) & = & O\left(X \right) U\left(X \right). 
\end{eqnarray}
$U(X)$ illuminates the object and the resulting scattered field propagates into the volume behind the target plane where its intensity is recorded at two different locations by the detectors, $P_1$ and $P_2$. These detectors record the intensity at a single point  and hence we ignore any spatial averaging effects caused by the finite size of the detector area \cite{Li:12, Kelly:13}. Let us first consider how to use the Fresnel transform to calculate the diffracted object field, $u_{z}(x)$, under normal illumination
\begin{eqnarray}
\label{S2.6}
u_z(x) & = &  FST_z \left\{O(X)  \right\}(x) \nonumber\\
u_z(x) & = & K_z \int_{-\infty}^\infty O(X)  \exp \left[\frac{  j \pi }{\lambda z}(x-X)^2 \right] dX, 
\end{eqnarray}
where $K_z = (1/\sqrt{j \lambda z})$, $\lambda$ the wavelength of the light, $z$ is the propagation distance. In practice we expect that $O(X)$ will have a finite extent. We can write this explicitly then in the following form $O(X)= T(X) p_L(X)$ and where $p_L(X)$ is an aperture function [defined in the same manner as Eq. (\ref{S2.2})] that defines the extent of the target in the object plane and where $T(X)$ describes the mathematical form of the target under examination. We now make use of the `modulation' property of the Fresnel transform. Given that
\begin{eqnarray}
\label{S2.6}
f_z(x) & = &  FST_z \left\{F(X) \right\}(x) \nonumber,
\end{eqnarray}
then the following relationship holds
\begin{eqnarray}
\label{S2.7}
FST_z \left\{F(X) \exp\left(j 2 \pi \rho X \right) \right\}(x)  & = & f_z(x-\lambda z \rho) \exp\left(j 2 \pi \rho x \right)  \exp\left(j\phi_{zc} \right), \nonumber\\
\end{eqnarray}
where $\phi_{zc}=-j \pi \rho^2 \lambda z$, \cite{Gori1981293,Stern:08b,Kelly:14}. With this result we can thus find a general expression for $FST_z \left\{ \tilde{O}(X)  \right\}(x)$,
\begin{eqnarray}
\label{S2.8}
\tilde{u}_z(x,\Delta_1) & = & \exp \left(j \phi_{1} \right) u_z(x-\lambda z f_{x1}) \exp\left(j 2 \pi f_{x1} x \right)  \exp\left(j\phi_{zc} \right) \nonumber\\
& = & \exp \left(j \phi_{1} \right) u_z\left(x- \frac{z}{f} \Delta_1 \right) \exp\left(j 2 \pi f_{x1} x \right)  \exp\left(j\phi_{zc} \right) 
\end{eqnarray}
We now rewrite Eq. (\ref{S2.8}) in the following manner
\begin{eqnarray}
\label{S2.9}
\tilde{u}_z(x,\Delta_1) & = & \exp \left(j \phi_{1} \right) a\left(x-\frac{z}{f} \Delta_1 \right) \exp\left[j \Theta \left(x-\frac{z}{f} \Delta_1 \right) \right],
\end{eqnarray}
where $a(x)= |u_z(x)|$ and where $\Theta(x)= \arg \left\{u_z(x) \exp\left(j 2 \pi f_{x1} x \right) \exp\left(j\phi_{zc} \right) \right\}$.
\\
\\
If we have an intensity detector situated at $x=x_1$, $z=z_1$, then as we physically shift our pinhole over the range $-D \leq \Delta_1 \leq D$, we will measure intensity values of $a^2(x)$ over the spatial range $(x_1-\frac{z}{f} D) \leq x \leq (x_1+\frac{z}{f} D)$. Hence we can scan the intensity of the diffracted field over the detector by varying the location of the pinhole in the diffuser-pair plane. Alternatively, we could move the detector position. We note that we would be unable to determine from intensity measurements alone whether the pinhole was being scanned over the diffuser-pair or whether the pinhole was stationary and the detector was moved. This is not true of the complex amplitude due to the various factors in Eq. (\ref{S2.8}). We shall return to this discussion in Section 3 and consider some implications for measuring the power of the object field. 
\subsection{Addition of a second pinhole}
We are now going to increase the complexity of the problem slightly by adding a second pinhole, situated at $R = \Delta_2$, to the pinhole mask. Since we are dealing with a linear system this means we can simple add the contribution from this second point source to the equations we have derived above. Let us refer to the complex amplitude at the location $P_1= (x_1, z_1)$  as 
\begin{eqnarray}
\label{S2.10}
u(P_1) & = & \tilde{u}_{z=z_1}(x_1,\Delta_1)  +\tilde{u}_{z=z_1}(x_1,\Delta_2) \nonumber\\
& = & \exp \left(j \phi_{1} \right) a\left(x_1-\frac{z_1}{f} \Delta_1 \right) \exp\left[j \Theta \left(x_1-\frac{z_1}{f} \Delta_1 \right) \right] \nonumber\\
&   & + \exp \left(j \phi_{2} \right) a\left(x_1-\frac{z_1}{f} \Delta_2 \right) \exp\left[j \Theta \left(x_1-\frac{z_1}{f} \Delta_2 \right) \right],
\end{eqnarray}
which we rewrite again for simplicity as 
\begin{eqnarray}
\label{S2.11}
u(P_1) & = & \exp \left(j \phi_{1} \right) a_1\exp(j \alpha_1)+ \exp \left(j \phi_{2} \right) a_2\exp(j \alpha_2).
\end{eqnarray}
In a similar manner we can write that 
\begin{eqnarray}
\label{S2.12}
u(P_2) & = & \exp \left(j \phi_{1} \right) b_1\exp(j \beta_1)+ \exp \left(j \phi_{2} \right) b_2\exp(j \beta_2).
\end{eqnarray}
\\
We are now in a position to examine in more detail the question of the intensities, $I(P_1)$ and $I(P_2)$, recorded by the detectors. We begin by noting the following cumbersome relationship
\begin{eqnarray}
\label{S2.13}
I(P_1)I(P_2) & = & u(P_1)u^*(P_1)u(P_2)u^*(P_2) \nonumber \\
& = & a_1^2 b_1^2 + a_2^2 b_1^2+ a_1^2 b_2^2+a_2^2 b_2^2\nonumber \\
& & + 2 a_1 a_2 b_1^2 \cos \left(\alpha_1-\alpha_2+\phi_1 -\phi_2\right) + 2 a_1 a_2 b_2^2 \cos \left(\alpha_1-\alpha_2+\phi_1 -\phi_2\right)\nonumber\\
& &  + 2 a_1^2 b_1 b_2 \cos \left(\beta_1-\beta_2+\phi_1 -\phi_2\right) + 2 a_2^2 b_1 b_2 \cos \left(\beta_1-\beta_2+\phi_1 -\phi_2\right) \nonumber\\
& & + 4 a_1 a_2 b_1 b_2 \cos\left(\alpha_1 - \alpha_2 +\phi_1- \phi_2 \right)  \cos\left(\beta_1 - \beta_2 +\phi_1- \phi_2 \right).
\end{eqnarray}
\subsection{Ensemble averaging}
Eq. (\ref{S2.13}) describes the product of the intensities located at $P_1$ and  $P_2$ for a given position of the diffuser-pair. If the diffuser-pair were displaced relative to each other by a significant amount (greater than the auto-correlation width of the surface function of the diffusers) we should expect the intensity values $I(P_1)$, $I(P_2)$ and their product  $I(P_1)I(P_2)$ to change, due to the fact that $\phi_1^n$ and $\phi_2^n$ are dependent on the relative diffuser-pair position. And each time one of the diffusers in the diffuser-pair is moved relative to the other, then $\phi_1^n$ and $\phi_2^n$ take on new and statistically independent values.  Using angled brackets to denote an ensemble average over a very large (strictly speaking an infinite) number relative diffuser-pair positions, it can be shown that 
\begin{eqnarray}
\label{S2.14}
\left< I(P_1)I(P_2)\right> & = & a_1^2 b_1^2 + a_2^2 b_1^2+ a_1^2 b_2^2+a_2^2 b_2^2\nonumber \\
& & + 2 a_1 a_2 b_1 b_2 \cos \left(\alpha_1-\alpha_2 - \beta_1+\beta_2   \right)
\end{eqnarray}
\\
We now wish to extend the result in Eq. (\ref{S2.14}) so that $N$ point sources in the diffuser-pair plane can be considered together. It can be shown that the general form for this relationship is given by  
\begin{eqnarray}
\label{S2.15}
\left< I(P_1)I(P_2)\right> =  DC + CT,
\end{eqnarray}
where
\begin{eqnarray}
\label{S2.16a}
DC = \left( \sum_{n=1}^N a^2_n\right) \left( \sum_{n=1}^N b^2_n\right),
\end{eqnarray}
and 
\begin{eqnarray}
\label{S2.16b}
CT = \sum_{m=1}^N \left[ \sum_{n=m+1}^N a_m a_n b_m b_n \cos \left(\alpha_m-\alpha_n - \beta_m+\beta_n   \right)  \right].
\end{eqnarray}
We would now like to extend the discrete analysis of $N$ contributing point sources so that a continuum of contributing point sources can be considered. Hence we shall rewrite Eq. (\ref{S2.16a}) and Eq. (\ref{S2.16b}) in integral form where we shall integrate over that area of the source plane that is allowed to contribute to the illumination of the object, i.e. we shall integrate over the ATM, 
\begin{eqnarray}
\label{S2.16-IntegralForm}
DC = \left[\int_{ATM} a^2 \left(x_1-\frac{z_1}{f} R_1 \right) dR_1 \right] \left[\int_{ATM} b^2 \left(x_2-\frac{z_2}{f} R_2 \right) dR_2 \right]
\end{eqnarray}
and 
\begin{eqnarray}
\label{S2.16b-IntegralForm}
CT & = & \int_{ATM}  \int_{R_1}^{ATM} a \left(x_1-\frac{z_1}{f} R_1   \right)  b \left(x_2-\frac{z_2}{f} R_1 \right) a \left(x_1-\frac{z_1}{f} R_2   \right)  \nonumber \\
& \times &  b \left(x_2-\frac{z_2}{f} R_2 \right) \cos \left[\alpha(R_1) - \alpha(R_2) - \beta(R_2)+\beta(R_1) \right] dR_2  dR_1 
\end{eqnarray}
This concludes our initial analysis of the experimental system depicted in Fig. \ref{CSpaperFigure2}.
\section{The correlation function}
We have just examined the intensity the detectors at $P_1$ and $P_2$ would measure. We would now like to consider some aspects of the power of the detected field and then to examine the definition of a correlation function that we will later use to identify specific objects. 
\subsection{Power illuminating the object}
We saw from Eq. (\ref{S2.2b}) that the total maximum power allowed into the optical system by the source is given by $2D$. This power will of course be reduced if source plane is stopped down with the ATM. Often when analyzing speckle systems the diffuser surface is assumed to be delta correlated, i.e. the auto-correlation of $V(R)$ is a Dirac delta function. This assumption greatly simplifies both speckle analysis and coherence theory, however it can lead to an unphysical result, see the discussion in Section 2 of Ref. \cite{Li:13b}. For example if the diffuser were in fact delta correlated, then the power of the source would be diffracted over an infinite extent in the illumination plane. This would imply that the maximum power $2D$ would be spread over an infinite extent with the result that very little power would in fact pass through the object plane with its finite extent of $2L$. And hence very little power would be measured at the detector locations, $P_1$ and $P_2$. 
\\
\\
However, in practice diffusers do not have a surface roughness function whose autocorrelation is delta correlated. Hence the light scattered by the diffuser only extends over a finite region in the object plane. There is a discrepancy between the theoretical model and the experimental situation. It has the following manageable implications. In the previous section we noted that each contributing point source had a unique and random phase value $\phi(R)$ associated with it. If the auto-correlation of $V(R)$ is not delta correlated then the same random phase value will make several weighted contributions to the ensemble averaging process. A good balance between the important simplifying theoretical assumption [that the source plane (diffuser) has a surface roughness function that is delta correlated or the extended light source has a delta correlated mutual coherence function] and ensuring that the actual physical experiment is correctly modelled would be to ensure that the average intensity at the object edges is approximately the same as the average intensity illuminating the center of the object, i.e. the delta correlated assumption should be accurate provided that:
\begin{eqnarray}
\label{R3.1}
\left< U(0)U^*(0) \right> =  \left< U(L)U^*(L) \right> = \left< U(-L)U^*(-L)\right>.
\end{eqnarray}
This will also ensure that a sufficient amount of power passes through the object plane and can be detected. Thus the power that then passes through the object plane is given by the following expression:
\begin{eqnarray}
\label{R3.2}
P_{tot} = \int_{-L}^{L} \tilde{O}(X) \tilde{O}^*(X) dX,
\end{eqnarray}
while the power of the field after the object when it is illuminated with only a single contributing point source and hence a single plane wave is given by
\begin{eqnarray}
\label{R3.3}
P_{obj} = \int_{-L}^{L} O(X) O^*(X) dX.
\end{eqnarray}
We will shortly use this power value, $P_{obj}$, to identify two different types of illumination conditions. First however we shall consider the extent of $u_z(x)$ in the measurement plane using power considerations.
\subsection{Space-bandwith product of the object}
We begin by making a general observation about $u_z(x)$ that is related to the spatial frequency content and spatial extent of the target, $O(X)$. We first note that Eq. (\ref{S2.6}) is a standard Fresnel diffraction problem that has been analyzed by many authors \cite{kelly:095801,kelly2011,Kelly:14}, however it is difficult to make any comments about the form of $u_z(x)$ without first making some assumptions about $O(X)$. We can say that $O(X)$ is limited in spatial extent by the aperture function $p_L(X)$. It then automatically follows that the spatial extent of $u_z(x)$ is in a strict mathematical sense infinite \cite{Gori1981293}. It also follows that the spatial frequency extent $O(X)$, as given by its Fourier transform, is infinite \cite{Gori1981293}. At this point it is useful to consider the Space-Bandwidth Product (SBP) of the signal $O(X)$ which can be used to define a region in phase space where a significant percentage of the signal's power resides \cite{Lohmann:93,Lohmann:96, Mendlovic:97, Mendlovic:97ex,LohmannBook2006,kelly2011,Kelly:14}. Defining the Fourier transform as 
\begin{eqnarray}
\label{S2.9b}
\bar{f}(v) & = & FT \left\{ f(X) \right\}  \nonumber \\
\bar{f}(v) & = & \int_{-\infty}^\infty f(X) \exp \left(- j 2 \pi v X \right)dX,
\end{eqnarray}
and the inverse transform as 
\begin{eqnarray}
\label{S2.9c}
f(X) & = & IFT \left\{ \bar{f}(X) \right\}  \nonumber \\
f(X) & = & \int_{-\infty}^\infty \bar{f}(v) \exp \left(+ j 2 \pi v X \right)dv,
\end{eqnarray}
it is possible to determine $\bar{O}(v)$ the Fourier transform of $O(X)$. We expect from Parseval's theorem \cite{BR1965} that the power in the spatial and Fourier domains to be conserved so that
\begin{eqnarray}
\label{S2.9d}
P_{obj}  =  \int_{-\infty}^{\infty} |\bar{O}(v)|^2 dv.
\end{eqnarray}
In practical cases however we are able to limit the extent of integration in the Fourier domain to a finite region, $-\Gamma \leq v \leq \Gamma$ such that    
\begin{eqnarray}
\label{S2.9d}
PR \times  P_{obj}  & = & \int_{-\Gamma}^{\Gamma} |\bar{O}(v)|^2 dv.
\end{eqnarray}
PR has a value close to arbitrarily close to unity. The product of $2L$ times the spatial frequency extent, $2 \Gamma$, is used to define the SBP of $O(X)$. Once this has been established we can also determine the effective Spatial Extent ($SE_z$) of $u_z$ and we now refer the reader to the following papers for more detail, \cite{Lohmann:93,Lohmann:96, Mendlovic:97, Mendlovic:97ex,LohmannBook2006,Kelly:14}.
\subsection{Power in the sequentially illuminated target}
In Section 2, we observed that for a given position $P_1$, it is possible to detect $a^2(x)$ over the spatial range $(x_1-\frac{z_1}{f} D) \leq x \leq (x_1+\frac{z_1}{f} D)$ by moving the location of the pinhole $\Delta_1$, see Fig. \ref{CSpaperFigure2}. Now we observe that if  $SE_z$ lies within the range $(x_1-\frac{z_1}{f} D) \leq x \leq (x_1+\frac{z_1}{f} D)$ then we can directly measure the total power of $u_z(x)$ by sequentially illuminating the target and summing the resulting detected intensities. This must give a total power of $P_{obj}$ which is calculated from Eq. (\ref{R3.3}). If we again simplify our notation and emphasize the sequential nature of this measurement, we would first  detect the following intensity for one contributing diffuser-pair point source
\begin{eqnarray}
\label{S2.9e}
\tilde{u}_z(x,\Delta_1) & = & \exp \left(j \phi_{1} \right) a_1\exp(\alpha_1),\nonumber\\
\end{eqnarray}
and 
\begin{eqnarray}
\label{S2.9f}
\tilde{u}_z(x,\Delta_2) & = & \exp \left(j \phi_{2} \right) a_2\exp(\alpha_2),\nonumber\\
\end{eqnarray}
and so on. Summing all of these together we get,
\begin{eqnarray}
\label{S2.9g}
P_{obj} = a_1^2+ a_2^2 + a_3^2 + ...
\end{eqnarray}
These observations also provide some insight into the interaction of speckle size and the target under illumination. If in fact, $SE_z > 2 D z /f$, then the power measurement we just described will result in a lower estimate for $P_{obj}$. For the measurement to be follow Eq. (\ref{S2.9g}) the K$\ddot{\mathrm{o}}$hler diameter must be sufficiently wide. We finally observe that if the pinhole mask were removed from the optical system, see Fig. \ref{CSpaperFigure2}, then our target would be illuminated with a speckle field, where the characteristic speckle size is given by $\lambda f/(2D)$, \cite{Good:1985,Li:12}.
\subsubsection{Illumination conditions: CPDR}
We can use the sequential illumination experiment just described to distinguish between two types of detection schemes. If a detector measures a power given by $P_{obj}$ then it is said to be located in a Complete Power Detection Region (CPDR). Within this scheme a correlation function that we shall define shortly, will always have a maximum value of unity when $P_1$ and $P_2$ are placed in the same location. 
\subsubsection{Illumination conditions: FPDR}
It is not always necessarily desirable to have this property in the detection scheme. Sometimes we can find out more information about a particular object or class of objects when they are illuminated with different ATM's as in Fig. 1 and Fig. 2. In this instance the detectors are placed in a Fractional Power Detection Region (FPDR) and hence $SE_z$ does not lie within the range $(x_1-\frac{z_1}{f} ATM) \leq x \leq (x_1+\frac{z_1}{f} ATM)$. The correlation function may sometimes obtain values greater than unity, an effect that is seen in some doubly scattering speckle phenomenon, \cite{ODonnell:82,Good:07,Li:13b}.  
\subsection{Defining the correlation function}
We begin by rewriting Eq. (\ref{S2.16b}) in a slightly different form:
\begin{eqnarray}
\label{S3.1}
CT & = & \sum_{m=1}^{N-1} \left[ \sum_{n=1}^{N-m} a_n a_{n+m} b_n b_{m+n} \cos \left(\alpha_n-\alpha_{n+m} - \beta_n+\beta_{m+n}   \right)  \right].
\end{eqnarray}
This form again involves a double summation, and we initially are going to concentrate on the summation that is in square brackets above. As $m$ increases the number of contributing elements in the second summation decreases.   We remember that the subscript on each of the elements of the summation refers to the location of the contributing point source in the diffuser-pair plane. When $m=1$ we can see that each of the contributing elements are neighboring points in the diffuser plane, and there are $N-1$ terms. When $m=2$ however the first point source is linked with the third point source, the second with the fourth and so on, such that there are $N-2$ terms. It perhaps easier to see this effect in the following illustrative table for five point sources, where the $\cos(\cdot)$ variable serves as a place-holder for the cosine terms from Eq. (\ref{S3.1}). \\
\\

\begin{tabular}{|c|c|c|c|c|c|} 
\hline 
$m$   & $n$=1 & $n$=2 & $n$=3 & $n$ = 4  & $n$ = 5  \\ 
\hline
1 & $a_1 a_2 b_1 b_2 \cos(\cdot)$  & $a_2 a_3 b_2 b_3 \cos(\cdot)$ & $a_3 a_4 b_3 b_4 \cos(\cdot)$ & $a_4 a_5 b_4 b_5 \cos(\cdot)$ & $a_5 a_6 b_5 b_6 \cos(\cdot)$   \\ 
\hline 
2 & 0  & $a_1 a_3 b_1 b_3 \cos(\cdot)$ & $a_2 a_4 b_2 b_4 \cos(\cdot)$ & $a_3 a_5 b_3 b_5 \cos(\cdot)$ &  $a_4 a_6 b_4 b_6 \cos(\cdot)$\\ 
\hline 
3 & 0  & 0 & $a_1 a_4 b_1 b_4 \cos(\cdot)$ & $a_2 a_5 b_2 b_5 \cos(\cdot)$ & $a_3 a_6 b_3 b_6 \cos(\cdot)$ \\ 
\hline 
4 & 0  & 0 & 0 & $a_1 a_5 b_1 b_5 \cos(\cdot)$ & $a_2 a_6 b_2 b_6 \cos(\cdot)$  \\ 
\hline 
5 & 0  & 0 & 0 & 0 & $a_1 a_6 b_1 b_6 \cos(\cdot)$  \\ 
\hline 
\end{tabular} 
\\
\\
\\
For a given value of $N$ contributing point sources it is possible to estimate the number of CT terms with the following formula: $(N-1)N+N/2$. The number of terms from the DC contribution in this case is $N^2$. Hence the ratio of DC terms to CT terms is given by $1- (2N)^{-1}$. In the limit as $N \to \infty$ then this ratio reduces to 1, indicating that 50\% of the total number of terms contribute to the DC component alone. The total number of terms is given by $2N^2-N/2$. The other terms then contribute to the CT in the following manner, see Fig. 3.
   \begin{figure}[htp!]
   \label{Figure3andB}
     \includegraphics[height=5cm]{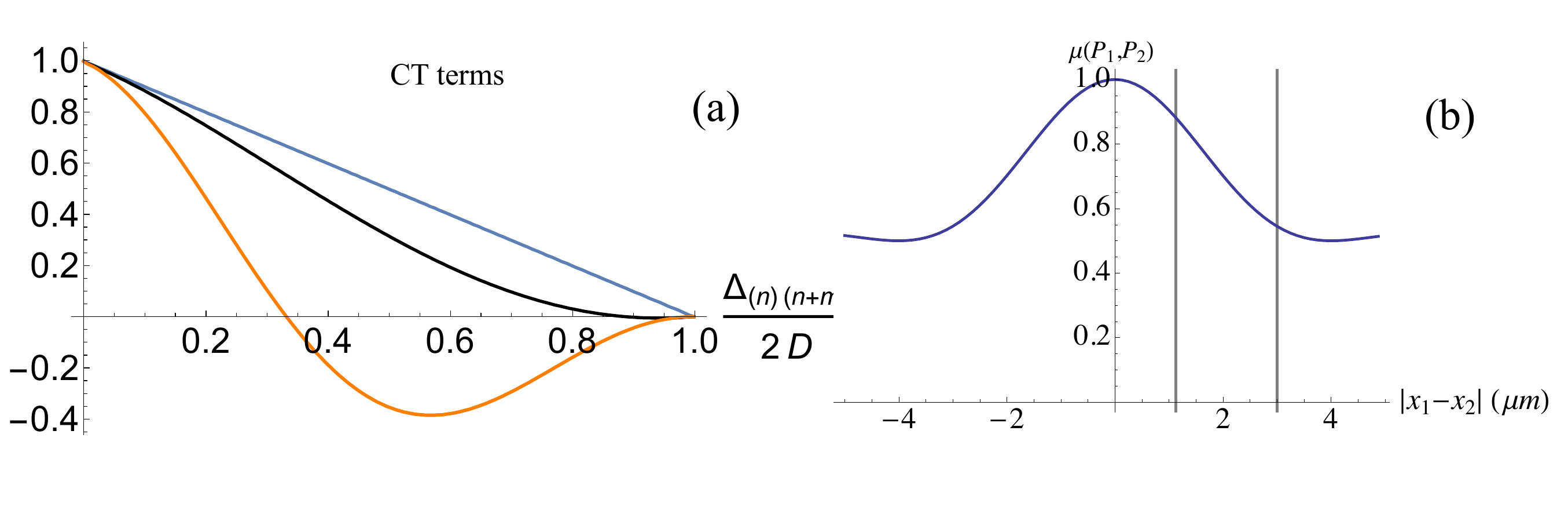}
     \caption{(a) The number of terms that contribute to CT are distributed in the following manner. There are more contributions from points that are located near to each other spatially. (b) Decorrelation of the coherence field as $P_2$ is moved away from $P_1$.}
    \end{figure}
We now choose an CPDR for the location $P_1$ using the experiment described in Section 2.A and examine Eq. (\ref{S3.1}) when $P_1$ and $P_2$ are actually located in the same place. When this happens the $a_n$ and $b_n$ terms are the same with the effect of canceling the cosine part of the cross-terms and Eq. (\ref{S3.1}) reduces to the following expression
\begin{eqnarray}
\label{S3.2}
\sum_{m=1}^{N-1} \left[ \sum_{n=1}^{N-m} a^2_n a^2_{n+m}    \right],
\end{eqnarray}
which if plotted as a function of spacing between contributing pinholes, exhibits the same type of linear decay as the figure plotted in Fig. \ref{CSpaperFigure2}.  If $P_1$ and $P_2$ are not in the same location then $a_n \neq b_n$ and the cosine term: $ \cos \left(\alpha_n-\alpha_{n+m} - \beta_n+\beta_{m+n}   \right) $ comes back into play. We thus choose the Normalization Factor (NF) to be 
\begin{eqnarray}
\label{s3.3}
NF = \sqrt{\left< I(P_1)I(P_1)\right> \left< I(P_2)I(P_2)\right>},
\end{eqnarray}
and now turn our attention to examining how $CT$ terms behave as $P_2$ is moved away from $P_1$ for an illustrative example.  In this simulation we assume the following parameters, for the optical system depicted in Fig. \ref{CSpaperFigure2}. We set $\lambda = 500$ nm, $D = 1.25$ mm, $f$ = 20 cm, $z_1 = z_2 = 20$ cm, and the target under illumination in the object plane is a simple rectangular aperture, $p_L(X)$, where $L = 2.5$ mm and $N$ = 600. We define a correlation coefficient in the following manner:
\begin{eqnarray}
\label{s3.4}
\mu \left(P_1,P_2\right) = \frac{\left< I(P_1)I(P_2)\right>}{NF},
\end{eqnarray}
and plot the results in Fig. 3 (b). It is clear that as $|x_1-x_2|$ increases, the decorrelation also increases. 
   \begin{figure}[htp!]
   \label{Figure5}
     \includegraphics[height=12cm]{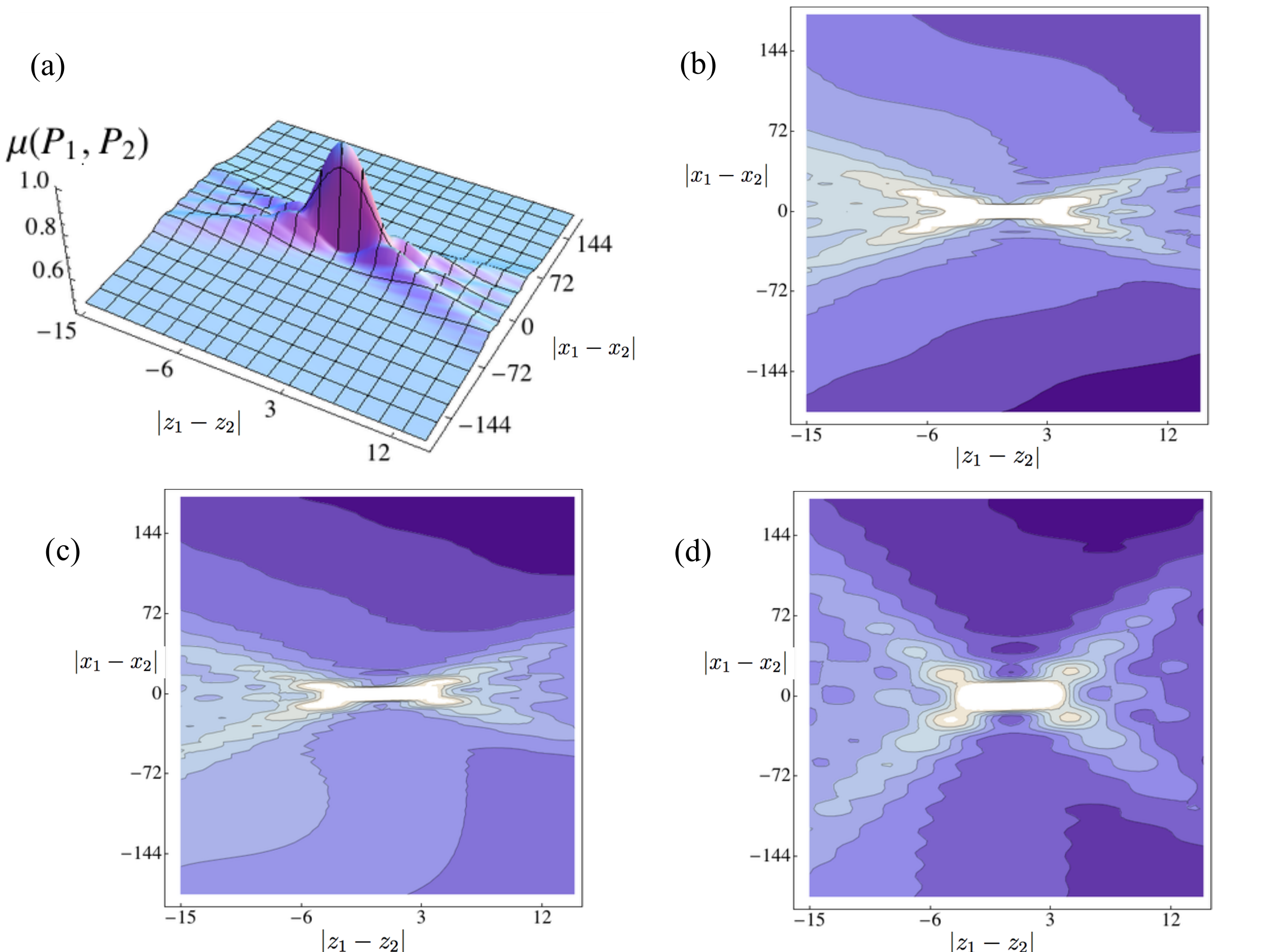}
     \caption{Decorrelation of the coherence field as $P_2$ is moved away from $P_1$, see text for details.}
    \end{figure}
In Fig. 3(b), we also examine the decorrelation for two spatial locations, $|x_1-x_2| = 1.125$ $\mu$m and $|x_1-x_2| = 3$ $\mu$m, and specifically plot the variation of the corresponding CT terms in Fig. 3(a) which are the black and orange curves respectively. As the $P_2$ and $P_1$ are separated spatially from each other, the cosine term: $ \cos \left(\alpha_n-\alpha_{n+m} - \beta_n+\beta_{m+n}   \right) $ starts to add more and more `out of phase' with each other with the effect that area under the black and the orange CT curves is lower than the maximum area that lies under the blue CT curve, i.e. when $P_1 = P_2$. Hence we understand that the expected amount of decorrelation is directly related to the ratio of the area under these two curves with a DC offset of 0.5. 
\\
\\
In order to better appreciate some of the properties of the coherence function, $\mu(P_1,P_2)$, we examine the 3-D distribution in Fig. 4 for a range of different detection locations. We choose $P_1$ and $P_2$ such that we operate in a CPDR mode. Again the optical system has the same parameters as those used to produce the plots in Fig. 3(b), however we now examine $\mu(P_1,P_2)$ for several different objects defined in the following manner:
\begin{eqnarray}
\label{s3.5}
O(X) = p_L(X) \left[ \exp\left( j 2 \pi f_{x1} X\right) + g \exp\left( j 2 \pi f_{x2} X\right) \right]
\end{eqnarray}
where $g$ = 0 or 1 and $f_{x1}$ and $f_{x2}$ are the spatial frequencies associated with the linear phase terms in Eq. (\ref{s3.5}). In Fig. 4(a) and (b), $f_{x1} = 1200$ lines/m and $P_1$ is located on axis with $x_1$ = 0 and $g = 0$. In Fig. 4 (a) we see that $\mu(0,0)$ has a value of unity. In Fig. 4(b) we present a contour plot version of the same data, and it can be seen that $\mu(x_1,x_2)$ is not symmetrical about the optical axis, and decreases as one moves to the bottom right of the plot. This is due to the spatial frequency component $f_{x1}$. In Fig. 4(c) we present the same experiment however now the location of $P_1$ has been changed and is now situated at $x_1 = 120$ $\mu$m. These properties can be used for alignment purposes. Finally in Fig. 4(d), we set $f_{x2}$ = $f_{x1}/2$, and leaving $x_1 = 120$ $\mu$m we can see `interference effects' of the overlapping spatial frequency components. 
\\
\\
Thus we conclude that if the target under investigation is known, then it is possible to calculate and predict the form of the correlation coefficient for any two detector pair positions, $P_1$ and $P_2$. It is also possible to modify the form of the correlation coefficient by changing the illumination of the source, for example by only illuminating with a specific ATM or by sequentially illuminating with different  ATM's. If the illumination region is `stopped down' then the detector signal, $\mu(P_1,P_2)$, can operate in a FPDR. As we shall see this effect can be used as another means of finding out more information about the object under investigation.
\subsection{Partial coherence, acquisition time and spectral width of source}
In the previous section we defined a correlation function $\mu(P_1,P_2)$. To measure this correlation function in an experiment we need to measure the following ensemble average: $\left<I(P_1)I(P_2) \right>$. This requires the following steps:
\begin{enumerate}
\item The intensity at $P_1$ and $P_2$ is recorded,
\item One of the diffusers in Fig. 2 is moved relative to the other diffuser and a new statistically independent illumination field is generated,
\item Again the intensity values at $P_1$ and $P_2$ are recorded,
\item Steps 1-3 above are then repeated for a very large (strictly speaking infinite) number of measurements. As we shall see in the following section only making a finite number of measurements results in an expected and quantifiable error.
\end{enumerate}
Since the diffusers are not moved during the detection of the light intensity, we define these experimental steps as being a `coherent' measurement. 
\\
\\
We now consider however what would happen if the diffusers were moved rapidly relative to each other and during the acquisition time (sometimes known as the integration time) $\Delta t$, of the intensity detector. If there were a very large number of statistically independent realizations of the illuminating field within the integration time of the electronic detector then the detected intensity at $P_1$ would converge to the average intensity value, i.e.  $\left< I(P_1)\right>$, and hence the correlation function, i.e. Eq. (\ref{s3.4}), would reduce to the following
\begin{eqnarray}
\label{IncohCorFunc}
\mu \left(P_1,P_2\right) = \frac{\left< I(P_1)\right> \left< I(P_2)\right>}{NF}.
\end{eqnarray}
Thus the averaging operation, indicated with the angled brackets, operates not on the product of the individual instances of intensity $I(P_1) I(P_2)$, but rather on the averaged intensities at the specific point detector locations. Since each detector would now only measure the average intensity value, this is the opposite extreme of the `coherent' measurement process and hence we define this situation to be an `incoherent' measurement. 
\\
\\
Now we observe that if there are several statistically independent illuminations of the object within the aquisition time of the detector that this is a partially coherent measurement. Hence this concept of coherence depends on the number of statistically independent measurements that are made within the aquisition time of the detector.
\\
\\
We would now like to extend our analysis so that a light source with a finite spectrum of wavelengths can also be considered. So we now examine what would happen if we replaced the diffuser plane pair and the monochromatic light source that serve to illuminate the sample in Fig. 2, and replaced them with an extended light source instead. This light source will have a particular bandwidth $\Delta \lambda$ that is related to a particular temporal spectral width $\Delta \omega$ by the following relationship for light in free space that $c = \lambda \omega$, where $c$ is the speed of light in a vacuum and $\omega$ is the temporal frequency of the light. 
\\
\\
Here we make a direct connection between a particular instance of our diffuser pair, which produces a random phase distribution $V(R)$ over the source plane and the instantaneous random (we also assume constant amplitude and random phase) distribution of the extended source. How rapidly will the instantaneous random phase distribution of the extended source change within the integration time of the detectors? We now make use of some of the results outlined by Goodman, see Chap. 5 of Ref. \cite{Good:1985} and also \cite{mandel1995optical}, and can define a `coherence time' $\tau_c$ for the light source. This `coherence time' is related to the form of the spectral profile, see for example Eq. (5.1-29) in Ref. \cite{Good:1985}. 
\\
\\
In order for us to make a `coherent' measurement, we require the speed of the optical detectors to be faster than $\tau_c$ in order to make a `coherent' measurement as defined above. Hence we have related the measurement technique in our simple optical system to a light source that is no longer mono-chromatic. Other techniques for analyzing and using partial coherence effects are discussed here \cite{Gopinathan:08b,Mehta:10,Rodrigo:14}.
\section{Tchebycheff's inequality, repeated trials and convergence.}
The equations that we have been examining, particularly Eq. (\ref{S2.13}) to Eq. (\ref{S3.1}), are valid for an ensemble averaging over a very large number of diffuser-pair positions. We remember that we can generate a new and statistically independent realization of a speckle for illumination, $U(X)$, by moving the diffusers relative to each other by a small amount (order of the auto-correlation of the surface profile), see Fig. \ref{CSpaperFigure2}. We now turn our attention to the convergence properties of this averaging process and ask: With what accuracy can we estimate the actual value of $\mu(P_1,P_2)$ when only a finite number, $M$, `realizations' are used in the averaging process? To address this question we turn to statistical methods to identify a `confidence interval', i.e. a percentage certainty that the estimated value lies within a specified narrow range of the actual value. We follow the analysis given in Section 8.2,  Chapter 8 of Ref. \cite{papoulis2002} and adopt the notation-style given there for random variables; where $\mathbf{\bar{y}}$ is a random variable and $y_i$ is a specific instance of this random variable. We now model the detection of $\mu(P_1,P_2)$ as a random process whereby we get the correct estimate plus a random error according to
\begin{eqnarray}
\label{S4.1}
\mathbf{\bar{y}} & = & \mu(P_1,P_2) + e,
\end{eqnarray}
where $\mathbf{\bar{y}}$ is a random variable with an unknown probability distribution. For a given measurement, $y_i$, it represents our estimate of the actual value, $\mu(P_1,P_2)$ plus a random error, $e$, which is related to the variance, $\sigma$, of $\mu(P_1,P_2)$ for different diffuser-pair realizations. We calculate our best estimate, $\mu_E(P_1,P_2)$, of the actual value $\mu(P_1,P_2)$ from a series of these different measurements; $[y_1 , y_2 , .... , y_M ]$ and set
\begin{eqnarray}
\label{S4.2}
\mu_E(P_1,P_2) = \mathrm{mean}\left\{ y \right\}
\end{eqnarray}
where `mean' performs an averaging operation. From \cite{papoulis2002} we set $e = \sigma/ \sqrt{M \tau}$ and using Tchebycheff's inequality, we find that 
\begin{eqnarray}
\label{S4.3}
\mathrm{Prob}\left \{ \mathbf{\bar{y}} - \frac{\sigma}{\sqrt{M \tau}} < \mu(P_1,P_2) < \mathbf{\bar{y}} + \frac{\sigma}{\sqrt{M \tau}}   \right\} > 1-\tau = \gamma
\end{eqnarray}
which shows that the exact $\gamma$ confidence interval of $\mu(P_1,P_2)$ is contained in the interval $\mathbf{\bar{y}} \pm  \sigma/\sqrt{M \tau}$. To use this relationship we need to know $\sigma$. While it should be possible to derive an analytical expression for the variance of $\mu(P_1,P_2)$ for finite number of realizations, it is  expected to be cumbersome (see Appendix E of \cite{Good:07}) and so we proceed in a more straight-forward manner. We make a series of intensity measurements at $P_1$ and $P_2$ and estimate the variance of $\mathbf{\bar{y}} $ from the sample variance 
\begin{eqnarray}
\label{S4.4}
s^2 =\left( \frac{1}{M-1} \right) \sum_{i=1}^M \left[y_i - \mathrm{mean} \left \{y \right \} \right]^2.
\end{eqnarray}
Eq. (\ref{S4.4}) represents an unbiased estimate of $\rho^2$ and tends to $\rho^2$ as $M \to \infty$, \cite{papoulis2002}. These results yield the approximate confidence interval of
\begin{eqnarray}
\label{S4.5}
\mu_E(P_1,P_2) - Z_{1-\tau/2} \frac{s}{\sqrt{M}}< \mu(P_1,P_2) <  \mu_E(P_1,P_2) + Z_{1-\tau/2}  \frac{s}{\sqrt{M}},
\end{eqnarray}
where 
\begin{eqnarray}
\label{S4.6}
u= \frac{1}{\sqrt{2 \pi}}  \int_{-\infty}^{Z_u} \exp  \left(-z^2/2 \right) dz.
\end{eqnarray}
If we assume a 95\% confidence interval, then $\tau$ = 0.05, and $Z_{0.975}  \approx 2$
\begin{eqnarray}
\label{S4.7}
\mu_E(P_1,P_2) =  \mu(P_1,P_2) \pm \epsilon,
\end{eqnarray}
where $\epsilon = 2 s / \sqrt{M}$. If we wish to determine a specific error range for our estimate (where we have a confidence of 95 \%) we find that we will need
\begin{eqnarray}
\label{S4.8}
M = \frac{4 s^2}{ \epsilon^2}
\end{eqnarray}
measurements.
   \begin{figure}[htp!]
   \label{Figure6}
     \includegraphics[height=12cm]{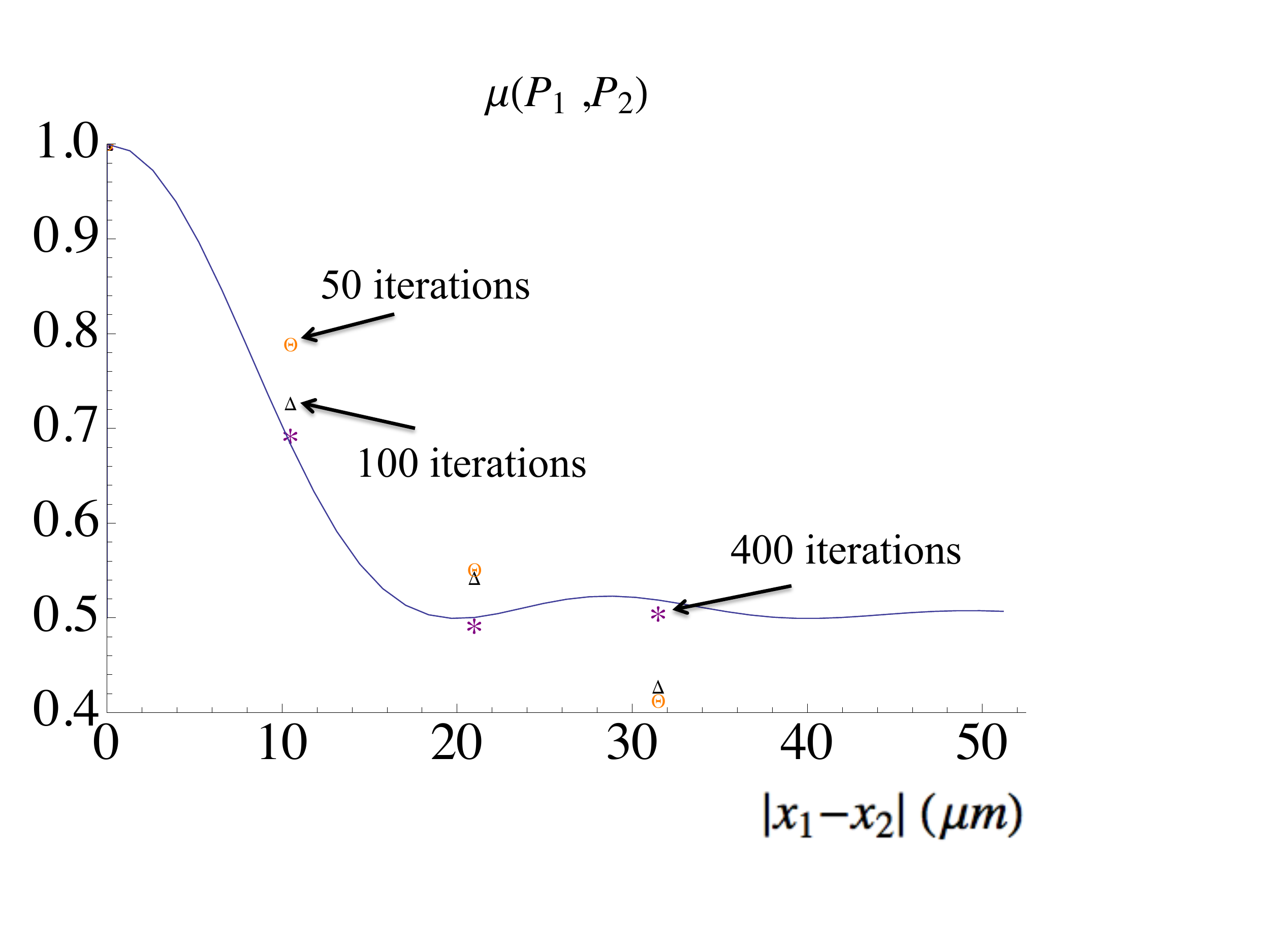}
     \caption{Process of convergence as the number of measurements is increased. As the number increases we can expect to achieve more certainty about the result in accordance with Eq. (\ref{S4.8}).}
    \end{figure}
\subsection{Repeated trials}
We now examine some of these statistical properties in more detail using a numerical simulation of the experiment depicted in Fig. \ref{CSpaperFigure2}. We do this with the following steps:
\begin{enumerate}
\item A random phase function is generated to describe the phase distribution of the light immediately after the diffuser-pair, $V(R)$.
\item This field is Fourier transformed to give a realization of an illumination field $U(X)$.
\item The illuminating field, $U(X)$ then multiplies the target function $O(X)$ and we then Fresnel transform the result to calculate the intensities at $P_1$ and $P_2$, \cite{Kelly:14}.
\item Repeat the steps $1 \to 3$, $M$ times.
\end{enumerate}
In this instance we use  simulation parameters similar to those used to generate Fig. 4. There are several differences however, we assume that the target we are examining is a converging lens, $\exp(-j \pi/ \lambda f X^2)$, where $f$ the focal length is 17 cm, with a diameter of 5 mm and we use 1600 samples to represent the target in the $X$ plane and $z_1$ = $z_2$ = 20 cm, $x_1$ = 0.
\\
\\
We expect that as more measurements are made that the accuracy of our estimate increases and this is confirmed in Fig. 5, where we present $\mu(P_1,P_2)$ and the estimates for three different lateral displacements of $P_2$ relative to $P_1$. In each case we also plot $\mu_E(P_1,P_2)$ for 50, 100 and 400 iterations respectively where each result is denoted with its own legend $\$$, $\#$, and $\ast$ respectively. Clearly as $M$ increases we approach the correctly calculated result, $\mu(P_1,P_2)$.
\section{Object identification}
\begin{figure}[htp!]
   \label{FIgFloat1}
     \includegraphics[height=9cm]{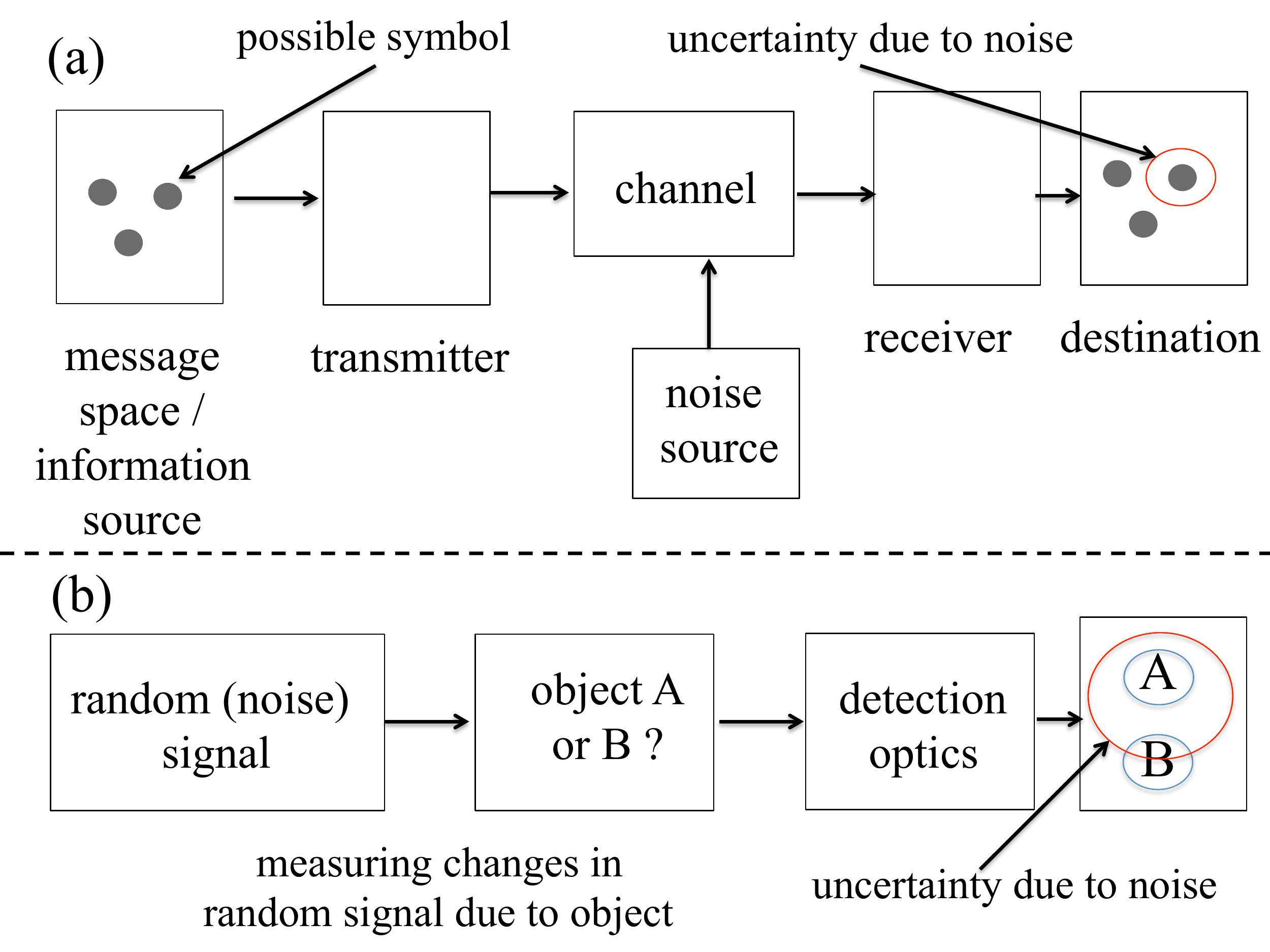}
     \caption{(a) Shannon's general communication channel. An information source (message space) produces a message for transmission through a noisy channel. After decoding there is uncertainty about the exact geometrical location of the message, (indicated by the diameter  of the red circle). Provided the circles surrounding the desired message are well separated, the correct symbol can be inferred. (b) Optical system we have in mind: A random signal is used to illuminate the object. The statistical properties of this signal are modified in a known way by a given object. By detecting these statistical deviations, different objects can be distinguished from each other. In this approach, we must illuminate the object with many different random signals. The ability to distinguish objects from each other improves as the  number of random input signals increases.}
    \end{figure}

In the previous sections we considered how to define our correlation function, and examined how it varied for different objects and established how this correlation function can be calculated. The theoretical derivation of this correlation function assumes that an infinite number of averages need to be made in order for the measurement process to converge. In Section 4, we specifically examined a means of estimating an error level when only a finite number of measurements are made. Hence we can state with a specific statistical certainty how accurate our experimental result would be for a given object and detection scheme. We now consider the problem of distinguishing different objects from each other and turn to some work from communication theory.
\\
\\
In the late forties, Shannon published two seminal papers \cite{shannon1948,SH1949} where he discussed communication systems and the transfer of information from a source to a destination over  a noisy communication channel. Specifically in \cite{SH1949} he envisages this process as consisting of several distinct parts: (i) an information source, (ii) a transmitter, (iii) the communication channel which modifies a signal in two distinct manners, one of which is due to a random noise source, (iv) a receiver, (v) and finally the end destination, which can be a person or a machine. He then proceeds to develop a geometrical model for the communication process. Each message produced by the information source is conceived of as being a single distinct geometrical point in N-dimensional space. In order to send such a `point' to the end recipient this message point is mapped to a physical signal that is suitable for transmission through the communication channel.  This mapping is very general and depends on the nature of the information being transmitted and the physical properties of the channel. Thus the function of the transmitter is to map the distinct message to an appropriate physical signal. At the receiver the detected signal is `un-mapped' and the relevant geometrical point and hence message can be determined by the recipient. This can be done unambiguously in a noise-free channel. In the presence of noise in the channel, this `un-mapping' operation (performed by the receiver) does not in general map to the same geometrical point as the `original' message but rather is mapped to a region of geometrical space about the correct message location. The greater the noise level,  the larger this region of uncertainty. Provided that all potential message points are well separated in N-dimensional space, then the correct message can be determined with near certainty, see Fig. \ref{FIgFloat1}. However if the noise is sufficiently great, it is possible that several potential messages could overlap with each other in N-dimensional space leading to uncertainty about what message was intended.
If the sender and the recipient agree in advance to limit the range of allowable messages or symbols that are to be sent over such a communication system, the job for the receiver is to correctly map the decoded signal to the intended message. 
\\
\\
Historically the Greeks are supposed to have developed communication systems based on lighting fires. These systems typically could send binary type messages, i.e. has Troy fallen or has it not? To send more general information requires a more complex set of symbols. Morse code maps the letters of the English language (symbols) to different series of `dots' and `dashes' and any type of detail can be transmitted at the expense of a more complicated coding and decoding process. These types of messages are discrete symbols. An information source can in principle produce a countably infinite number of these discrete symbols and each symbol will require its own unique identifying signature so that it can be unambiguously identified.
\subsection{Sets of discrete objects}
We highlight the main idea of object detection that we pursue here in Fig. \ref{FIgFloat1} (b). We imagine that we interrogate an unknown symbol (object A or B?) with a set of random signals. The symbol interacts with these random signals modifying them statistically in a  manner that is unique to each symbol. Hence by measuring the statistical changes to the interrogating random signals, the pertinent symbol can be identified with an arbitrarily high certainty. With a finite number of measurements there will be an uncertainty about identifying the correct object, as indicated in Fig. 6(b), where object A is recognized as being the most likely symbol. By making further measurements the diameter of the red circle (around symbol A) can be made smaller and our certainty about the object can be increased in a manner analogous to Shannon's geometric interpretation of a communication system.  Our level of statistical certainty was discussed in Section 4.
\\
\\
We can therefore now imagine the following situation. We have two different objects; one as before a converging lens of focal length, $f = 17$ cm with lens diameter of 5 mm, and the second a cosine grating with a spatial frequency, $\Gamma$ = 600 lines/m and diameter 5 mm,
\begin{eqnarray}
\label{S4.8}
O_A(X) = \exp \left(\frac{- j \pi X^2}{\lambda f} \right) p_L(X)
\end{eqnarray}
and
\begin{eqnarray}
\label{S4.9}
O_B(X) = \cos \left(2 \pi \Gamma X    \right) p_L(X).
\end{eqnarray}
Using the equations; Eq. (\ref{S2.15}) to Eq. (\ref{S3.1}) and Eq. (\ref{s3.4}), we calculate that 
\\
\\
\begin{tabular}{|c|c|c|c|c|c|} 
\hline 
$ $   & Object A & Object B   \\ 
\hline
$ $ & $\mu_A(P_1,P_2) \approx 0.498$ & $\mu_B(P_1,P_2) \approx 0.614$  \\ 
\hline 
\end{tabular} 
\\
\\ 
\\
when $x_1$ = 0, $x_2 = 120.75$ $\mu$m and $z_1 = z_2 = $ 20 cm. It is possible to calculate in advance the coherence field for each object which we refer to as $\mu_A(P_1,P_2)$ and $\mu_B(P_1,P_2)$ respectively, which is plotted in Fig. 6 (a) as black and orange plots respectively. We can see that both plots produce quite similar distributions when $|x_1 - x_2| \approx 0$, however there is a significant difference when $x_1 =0$, and $x_2 \approx 120$ and accordingly this is where we choose to place our detectors. Using again 1600 samples to represent the object in the sample plane we implement a 3500 numerical simulations to mimic the experimental measurement technique. The calculated values indicate that $\mu_E(P_1,P_2)$ = 0.5927 and have an estimated variance $s^2 \approx $ 1.3. Hence we can be 95\% certain that the actual value for lies within a range $\epsilon $ = 0.05 of the estimated value $\mu_E(P_1,P_2)$, as indicated in Fig. 6 (b). This indicates with a strong statistical likelihood that Object B is in fact the actual object under examination. This particular test has been carried out in an CPDR region.
   \begin{figure}[htp!]
   \label{Figure7}
     \includegraphics[height=12cm]{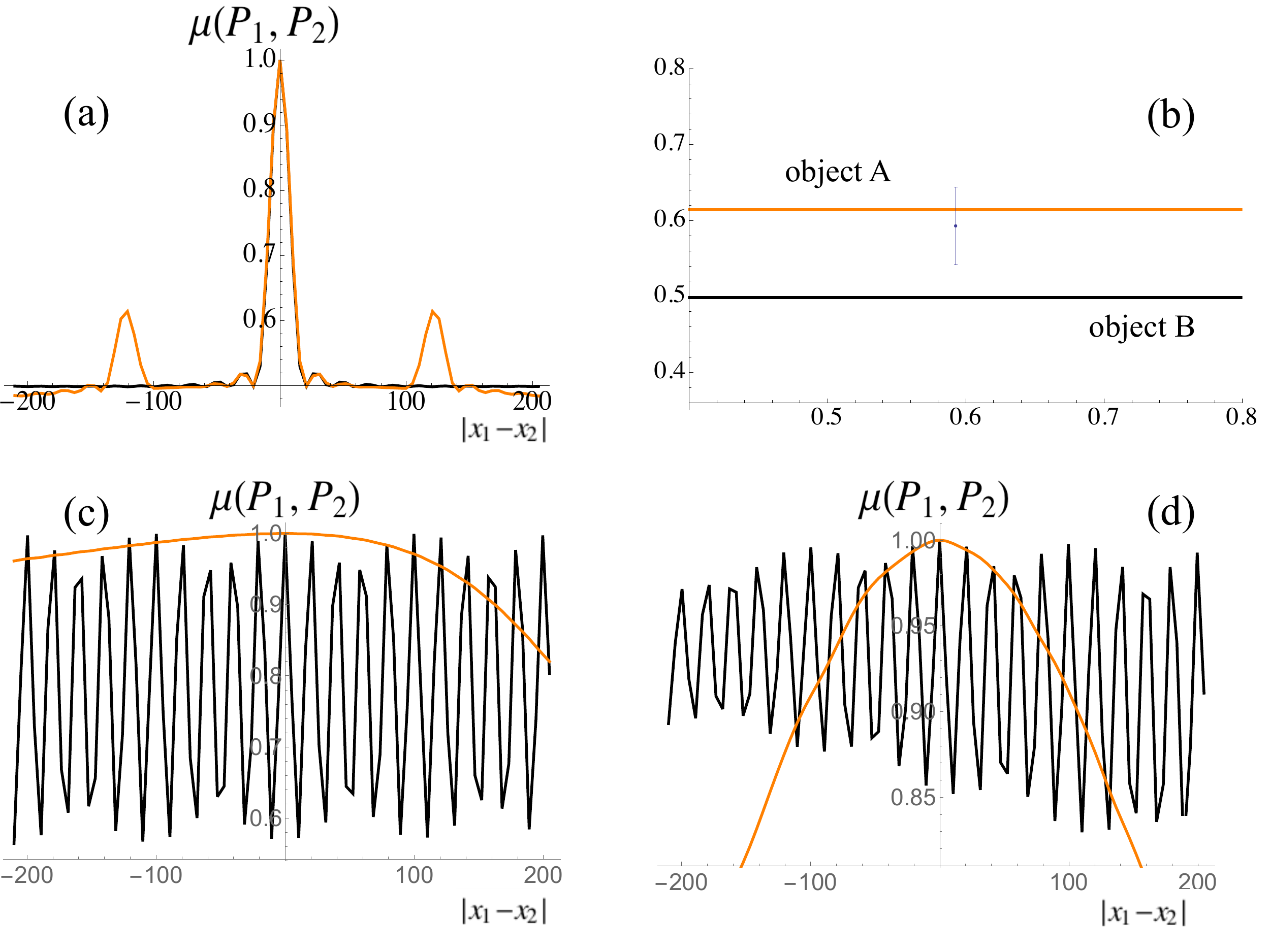}
     \caption{We present several figures showing how $\mu(P_1,P_2)$ changes for Object A (in black) and B (in orange) ( under differing illumination conditions. In (a), we are operating in CPDR, while in (c) and (d) we are in FPDR regime. In (b), we present the statistical results of the numerical simulation, where we are 95\% certain that the correct value lies within error range indicated. Hence we can correctly distinguish object B from object A. } 
    \end{figure}
\\
\\
As we begin to add more symbols to the original set of two objects, it is possible that more than one symbol will have the same value for $\mu(P_1,P_2)$ and hence it would seem to be impossible to distinguish between them. In fact we can see from Fig. 6 (a) that both objects have very similar distributions about $x_1$ = 0. It is possible to overcome this type of problem by (i) Using different locations for $P_1$ and $P_2$ or (ii) Using the detection scheme in FPDR region instead of CPDR.  This is indicated in Fig. 6 (c) and (d) where we test the same objects under different illumination conditions,  where an aperture is used to `stop down' the regions in the diffuser plane which contribute to the illuminating light, (c) 250 $\mu$m aperture located at $R = -2.42$ mm and (d) the same aperture located at $R$ = 0.83 mm. As can be seen this changes the characteristic signature in both plots quite significantly. Indeed underlines the importance of the illumination conditions in the detection scheme and should be considered as part of the encoding and decoding process. If each symbol has a unique identification signature (detected signal) and hence a unique geometrical location in N-dimensional space, then each symbol can be decoded uniquely from each other by using an array of different point locations $P_1$ and $P_2$. The appropriate encoding and decoding schemes depend on the particular optical application. 
\\
\\
Hence we conclude that if we have a finite number of objects and associate each object with its own statistical signature through a combination of detector locations and illumination patterns, (such as using different ATMs) then it will be possible to distinguish between them by making a known number of measurements and within a known statistical certainty.
\subsection{`Continuous' Objects}
There are however another class of possible messages. In Shannon's analysis he notes that some messages are in fact continuous signals like a short-wave radio broadcast or an analog television stream. In Section II and III of his paper \cite{SH1949} he shows how continuous signals can be represented using a finite number of samples and recovered using ideal reconstruction filters using the sampling theorem and thereby defines a space-bandwidth-product. Once the message has been translated into a finite number of samples it can again be interpreted as being a location in N-dimensional geometric space. Slepian writes ``\emph{Shannon himself was unhappy with his method of bridging the gap from the time-discrete to the time-continuous case. Indeed, it was as a result of questions he raised in trying to make rigorous this notion of 2WT degrees of freedom for signals of duration $T$ and bandwidth $W$ that the research leading to the Landau-Pollak theorem got under way}.'' \cite{Slepian-On-Bandwidth}. 
\\
\\
In 1969, Toraldo di Francia considers the degrees of freedom in an image which is related to the space-bandwidth product of an optical signal \cite{DIFRANCIA:69}. He uses ideas from information theory and applies them to optics which was also developed by Gabor, Slepian \cite{Slepian-On-Bandwidth}, Lohmann \cite{LohmannBook2006,Lohmann:96} and others \cite{Mendlovic:97,Mendlovic:97ex,Zalevsky:00,Piestun:00}. He notes that practically speaking it is possible for many different complex objects to produce identical intensity images. This can produce difficulties if we attempt to distinguish between different complex objects (optical signals) using intensity measurements. Hence it is important that if we have a set of discrete complex objects or equivalently symbols, that each object should have its own unique identification signature (its own unique space-spatial-frequency distribution or individual basis set representation \cite{Slepian-Pollak-1961}) and hence a unique location in N-dimensional geometric space.  Under these conditions it is possible to distinguish a single object from a given set of objects or symbols, after a suitable number of intensity measurements are made and with a specific certainty that was identified in Section 4. 
\\
\\
As we have noted it is possible to represent a continuously varying signal with a finite number of terms provided that specific conditions, for example the Nyquist sampling theorem, is fulfilled. Shannon recognized that this allowed him to represent continuous signals like an analog television broadcast in terms of a space bandwidth product and hence a discrete location in N-dimensional geometric space. We will try to extend this so that continuously varying objects, for example a lens with an unknown focal length or a complex phase grating with several unknown spatial frequencies, can be identified from experimental measurements of $\mu(P_1,P_2)$. Using concepts from optical information theory we will write the object under investigation using an orthogonal basis set expansion with a finite number of weighting terms. As Toraldo di Francia notes all physically realizable optical signals can be effectively represented with a finite number of weighted basis set terms, which is theoretically underpinned with the work from Slepian and Pollak \cite{Slepian-Pollak-1961}. Once we represent the unknown object in this manner, we proceed to make a series of experimental measurements. The task then is to vary the weights of the contributing basis set terms so as to minimize in a least squares sense the error between the measured $\mu_E(P_1,P_2)$ and $\mu(P_1,P_2)$. We begin by writing 
\begin{eqnarray}
\label{S4.10}
O(X) = \sum_{k=1}^K \psi_k \Omega_k(X),
\end{eqnarray}
where $\psi_k$ are the weights of each orthogonal basis set contribution, $\Omega_k(X)$. We now attempt to minimize the error 
\begin{eqnarray}
\label{S4.10}
E(\overline{\mathrm{P}},\overline{\mathrm{\psi}}) = \sum_{q=1}^{Q} \left[\mu_E(\overline{\mathrm{P}})- \mu(\overline{\mathrm{P}},\overline{\mathrm{\psi}}) \right]^2,
\end{eqnarray}
where we have introduced the vector $\overline{\mathrm{P}}$ representing a finite number $Q$ of difference locations $P_1$ and $P_2$, or a set of $Q$ different measurement locations for the detectors $P_1$ and $P_2$. The term $\overline{\mathrm{\psi}}$ represents the vector of weights for the basis set representation of $O(X)$ which are to be adjusted so as to minimize $E$, i.e. 
\begin{eqnarray}
\label{S4.11}
\overline{\mathrm{\psi}} = \left[\psi_1, \psi_2 ... \psi_K \right].
\end{eqnarray}
If $K> Q$ then the problem is ill-posed, there are more unknowns than measurements and consequently many solutions. We now minimize the error with respect to the weights of the basis set representation by solving the following equation 
\begin{eqnarray}
\label{S4.12}
\sum_{k=1}^K \frac{\partial E}{\partial \psi_k} = 0.
\end{eqnarray}
If Eq. (\ref{S4.12}) can be expressed in a linear manner then it is possible to find a unique solution. In the general case however it is expected that Eq. (\ref{S4.12}) is a non-linear expression and hence is best solved using iterative non-linear least mean squares techniques such as Newton-Raphson or Levenberg-Marquardt. A non-linear expression will generally have multiple solutions and hence we need an initial guess as what $\overline{\mathrm{\psi}}$ is to start the iterative process. The non-linear solution will converge to the nearest minimum. 
\\
\\
Comparing Shannon's approach to that just outlined,  several important consequences emerge. Once the sampling theorem is obeyed then a unique solution or message in N-dimensional geometric space is unambiguously defined. In the situation described here the results are not as straight-forward. For example, if Eq. (\ref{S4.12}) is non-linear then it is possible that there are multiple minima and hence multiple possible messages - the solution is not necessarily unique. Hence here in addition to representing the object using a finite number of weighted basis terms, we are also required to have a good initial guess at the starting vector $\overline{\mathrm{\psi}}$ so that we converge quickly to the correct minimum. Implementing such an approach in a practical scenario requires carefully choosing the illumination conditions (CPDR and FPDR), the detection locations $\overline{\mathrm{P}}$ and an appropriate basis set. Some a-priori knowledge about a good initial guess at $\overline{\mathrm{\psi}}$ will help significantly with the convergence process. These factors depend very much on the class of objects under examination and hence are not pursued further in this manuscript.
\section{Wavelength and polarization encoding}
Until now we have only considered light that has a single wavelength and that is linearly polarized. The results and conclusions we have derived in the preceding sections can however be extended relatively easily to include both polarization effects and different colors of light. Under the current assumptions, i. e. that both the paraxial approximation and TEA are valid, we need only modify $O(X)$ so that it explicitly depends on both $\lambda$ and the polarization state. These extra degrees of freedom can be used to provide significantly more information about the object under inspection. In this instance a particular wavelength of light is chosen and the coherence measurement technique outlined in Sections 2-5 is applied to find out information about the object. Then the wavelength is changed and the process is repeated. It is possible perform this measurement for multiple wavelengths simultaneously, however a more sophisticated detection scheme would be necessary, perhaps with color filters over different arrays of PIDs. 
\\
\\
We remember that in this analysis we have always assumed that the scalar approximation is valid. Therefore we may consider different polariziation components independently of each other. Using a quarter-wave plate in Fig. 2 it is possible to turn linearly polarized plane wave light into circularly polarized light which is then used to illuminate the object with two different polarizations simultaneously. If the object reacts differently to different polarizations then the object will have two different transmittance functions which can be treated separately from each other in line with our assumption that a scalar model is accurate. Orthogonal polarizations can be measured using PID and a polariziation filter.
\section{Discussion and conclusion}
In this manuscript an alternative approach to object identification has been undertaken. We have emphasized ideas from communication theory, in particular Shannon's work on communication over a noisy channel. In his work he imagines that all possible symbols produced by an information source are represented as distinct points in N-dimensional geometrical space. Here, we imagine that the objects we wish to distinguish from each other are also known in advance and form characteristic and unique signatures. In our case the unknown object is illuminated sequentially with a finite number of random fields, producing two series of random intensities that are measured at two different spatial locations by two point intensity detectors. The statistical relationship between these intensity series recorded in the 3-D volume behind the object can be measured and should follow a specific statistical distribution that depends on both the object and the illumination conditions. For a given discrete set of objects, each must have a unique distribution so that it can be unambiguously determined with a finite number of measurements.
\\
\\
The results presented here are exact provided that several different approximations are valid. The paraxial scalar approximation and the `Thin Element Approximation' (TEA) are assumed and that the intensities measured by the detectors are approximately constant over the light sensitive area of the detectors so that any spatial averaging effects are negligible. We also assume that the K$\ddot{\mathrm{o}}$hler lens in the system is not only thin but infinite in extent, which is clearly not physically true. The accuracy of this last assumption is surprisingly accurate \cite{KellyFiniteAp3D,Kelly:07b}. In principle, the accuracy of these assumptions can be tested by choosing an exact known object such as a square or circular opening. The resulting statistical distribution should fall within the predicted range within specific statistical limits. A set of measurements that does not correspond to the presumed situation must then raise questions about the assumptions that are made within a specific statistical limit that is discussed more explicitly in Section 4.
\\
\\
In this paper the bulk of the analysis has been done for a specific wavelength and polarization, however as we saw in Section 6 this naturally extends different wavelengths and polarizations. In fact the wavelength and polarization provide more degrees of freedom to us and allow us to consider different physical properties of the object - the better to distinguish it from the other potential objects in the set (a set that may contain an infinitely large number of discrete objects but is countable). By limiting ourselves to a finite set of discrete objects, each with its own unique statistical signature, we can define with an arbitrarily high certainty our `confidence level' about the actual object under test. This type of approach also lends itself to the `compressive sensing' paradigm see for example the following references: \cite{SternRivenson2011}
\\
\\
A significant limitation of this technique is what happens when we have no a-priori information about the object we are examining. Or if the set of objects can be described with functions whose parameters vary continuously. It is possible in this case that many different objects have identical statistical signatures as Toraldo di Francia noted about images and objects \cite{DIFRANCIA:69}, or the difficulties that Shannon noted and Slepian et. al. addressed when moving from the ``time-discrete to time-continuous'' case. This is by no means a trivial problem, it depends on our state of knowledge of the objects in our discrete set, and the basis set we use to define the object class. If we are in complete ignorance about the object scene under investigation then a traditional imaging system is the best approach to revealing what is before us. We note however that investigating objects or the time-behaviour of samples does not necessarily have to be performed in an imaging environment, although that seems most intuitive to us \cite{Marr187}. It is conceivable that we might be able to interpret understand objects by first predicting what we might expect to detect along the lines outlined here \cite{CohenJoel}. We also note however it is possible to use both an imaging system in conjunction with the detection scheme described here. 
\section*{ACKNOWLEDGMENTS}
DPK is now Junior-Stiftungsprofessor of ÒOptics DesignÓ and is supported by funding from the Carl-Zeiss-Stiftung, (FKZ: 21-0563-2.8/121/1).

\end{document}